\documentclass[aps,pra,twocolumn,superscriptaddress,10pt,longbibliography]{revtex4-2} 
\usepackage{graphicx}  
\usepackage{dcolumn}   
\usepackage{bm}        
\usepackage{amssymb}   
\usepackage{amsmath}   
\usepackage{braket}    
\usepackage{mathtools}
\usepackage[linktocpage]{hyperref} 
\usepackage{cleveref}
\usepackage{subfig}
\usepackage{float}
\usepackage{xcolor}

\def\black#1{{\color{black} #1}}

\usepackage[colorinlistoftodos]{todonotes} 

\usepackage{fancyhdr}
\newcommand{\lab}[2]{\tag{#1.#2}\label{eqn:Sec#1_Eqn#2}}
\newcommand{\reff}[2]{\ref{eqn:Sec#1_Eqn#2}}

\begin{document}

\fancyhead[R]{\ifnum\value{page}<2\relax\else\thepage\fi}

\title{The enduring relevance of the Jaynes-Cummings model: a personal perspective}
\author{Peter L. Knight$^{1}$, Christopher C. Gerry$^{2}$, Richard J. Birrittella$^{3}$ and Paul M. Alsing$^{4}$ \\
\textit{\textit{$^{1}$Blackett Laboratory, Imperial College, London SW72AZ, UK}
	\\
\textit{$^{2}$Department of Physics and Astronomy, Lehman College,
The City \\University of New York, Bronx, New York, 10468-1589,USA\\}
\textit{$^{3}$Booz Allen Hamilton, 8283 Greensboro Drive, McLean, VA 22102, USA\\} 
\textit{$^{4}$University at Albany-SUNY, Albany, NY 12222, USA} 
}}

\date{\today}

\begin{abstract}
In this short perspective article we present our personal highlights on how the Jaynes-Cummings model has become a central model to describe spin-boson couplings underpinning much of modern quantum optics. To the current authors, the key contribution is a demonstration of a measurable effect that showed the discreteness of the quantized radiation field.
\end{abstract}
\pacs{}

\maketitle

\thispagestyle{fancy}
\section{\label{sec:EarlyHistory} Early History}

We discuss how the Jaynes-Cummings model \cite{ref:JaynesOG} has become the basic building block of resonance physics, demonstrating the non-perturbative time-dependent oscillation of transition probabilities when a single quantized field mode is coupled to a two-level system. It underpins much of quantum optics and quantum information processing \cite{ref:GerryBook} and 61 years after the original publication continues to be a highly cited paper.

Atom-field resonant excitation has a long history \cite{ref:Rabi1} (and see \cite{ref:GerryBook} and references therein).  Rabi developed the semi-classical resonance model where the field is described classically to describe atomic beam resonance experiments.  An essential assumption made was the rotating wave approximation (RWA) \cite{ref:GerryBook} \footnote{Going beyond the RWA has been an unwise theoretical obsession hardly ever justified if the two-level approximation is employed: counter-rotating non-resonant terms are always going to couple to other atomic levels, unless dealing with a genuine spin with no other nearby levels.}, justified by the closeness of a resonance and the lack of other nearby atomic levels.  This work built on earlier time-dependent work on non-adiabatic Majorana ``flops" \cite{ref:RamseyBook}.  Indeed, non-adiabatic transitions were studied even before Rabi's resonance model.  The first suggestion came from C. Darwin (\textit{the} Darwin's grandson!) in 1927 \cite{ref:Darwin} and studied experimentally by Phipps and Stern and theoretically by Guttinger and very thoroughly by Majorana \cite{ref:Majorana} and confirmed experimentally by Frisch and Segre (see \cite{ref:RamseyBook} and references therein); hence these were referred to by the atomic beams community as ``Majorana flops" as they created spin flips that ``flopped" the focused atomic beam back to the detector.  Rabi, aware of these developments, then built on them to construct his radio frequency resonance approach a few years later. 

Jaynes, in a famous Stanford Microwave Lab report \cite{ref:StanfordMicrowaveReport}, extended the semi-classical approach to describe maser action and shortly after brought in F.W. Cummings, his then-graduate student, to analyze a fully quantized version of the Rabi model assuming Fock (number) states of the field, and the resultant paper is the now-famous Jaynes-Cummings Model (JCM) \cite{ref:JaynesOG}.

We denote the atomic resonance frequency and cavity frequency as $\omega_{0}$ and $\omega$, respectively, with Pauli raising and lowering operators to describe absorption and emission, with an atom-field coupling strength denoted by $\lambda$.  In the RWA, the Jaynes-Cummings Hamiltonian is given by 

\begin{equation}
    \hat{H} = \frac{1}{2}\hbar\omega_{0}\hat{\sigma}_{3} + \hbar\omega\hat{a}^\dagger\hat{a}+\hbar\lambda\left(\hat{\sigma}_{+}\hat{\hat{a}}+\hat{\sigma}_{-}\hat{a}^\dagger\right).
    \lab{1}{1}
\end{equation}

The JCM is ubiquitous in quantum optics as the fundamental spin-boson model for a single photon interacting with a single two-level atom (in the RWA) inside a cavity, spawning the field of cavity quantum electrodynamics (QED, see e.g. Haroche and Raimond \cite{ref:HarocheBook} and references therein).   It's investigations and extensions in the field of quantum optics have been extensively explored. 
\black{A number of early papers investigated the interaction of a two level system with a single quantized field mode including those by Paul \cite{ref:Paul:1963} and Frahm\cite{ref:Frahm:1966}.}

Returning to the JCM Hamiltonian of Eq.~(\reff{1}{1}), the probability that the system remains in the initial state 
\black{
$\ket{i}=\ket{e}_a\otimes\ket{n}_f$ 
(atom in excited state $\ket{e}_a$ and field in Fock state $\ket{n}_f$ containing $n$ photons)
} is 

\begin{equation}
    P_{i}(t) = |C_{i}(t)|^{2} = \cos^{2}\left(\lambda t\sqrt{n+1}\right),
    \lab{1}{2}
\end{equation}

\noindent while the probability that it makes a transition to the state $\ket{f}$ is 

\begin{equation}
    P_{f}(t) = 1 - P_{i}(t) =\sin^{2}\left(\lambda t\sqrt{n+1}\right).
    \lab{1}{3}
\end{equation}

\noindent Cummings, in a later paper \cite{ref:Jaynes5}, addressed the role of photon statistics in the JCM and showed how the distribution of photon numbers leads to a spread of oscillatory Rabi frequencies and an inevitable dephasing of the otherwise regular sinusoidal evolution.  For the most classical of the pure quantized field states, a coherent state of amplitude $\alpha$, given by

\begin{equation}
    \ket{\alpha} = e^{-\tfrac{1}{2}|\alpha|^{2}}\sum_{n=0}^\infty\frac{\alpha^{n}}{\sqrt{n!}}\ket{n},
    \lab{1}{4}
\end{equation}

\noindent with average photon number $\bar{n}= |\alpha|^{2}$, we have

\begin{equation}
    P_{n} = |\braket{n|\alpha}|^{2}=e^{-|\alpha|^{2}}\frac{|\alpha|^{2n}}{n!} = e^{-\bar{n}}\frac{\bar{n}^{n}}{n!}.
    \lab{1}{5}
\end{equation}

\noindent Thus the difference in occupation probabilities between the initial and final states, known as the atomic inversion and denoted $W(t)$ is 

\begin{equation}
    W(t)=  \braket{{\hat{\sigma}}_{3}} = e^{-\bar{n}}\sum_{n=0}^{\infty}\frac{\bar{n}^{n}}{n!}\cos\left(2\lambda t\sqrt{n+1}\right).
    \lab{1}{6}
\end{equation}

This leads to a dephasing of the Rabi oscillations, now known as a Cummings collapse, due to the spread of photon numbers.  This is a unique quantum feature that is important to note as even a ``classical" state (coherent or thermal) has such a spread: the evolution is photon-number sensitive in a truly non-classical manner.

The collapse time found by Cummings is governed by the inverse of the one-photon coupling strength $t_{c}=(2\lambda)^{-1}$.  But this collapse is non-dissipative$-$ what happens if one pursues the evolution for longer times? In hindsight, it is clear that we should expect some kind of return of coherent oscillations given the discreteness of photon occupation number.  The literature does contain precursors of what turned out to be revivals (Meystre \textit{et al.} \cite{ref:Meystre}, Stenholm \cite{ref:Stenholm}, etc.) in numerical studies of the JCM albeit with truncated numerics at this early stage in computational analysis in the early 1970s.  

\section{\label{sec:Section2} Collapse and Revivals}

The systematic study of the effects of photon statistics on the long-term time evolution of the JCM was pioneered by J. H. Eberly and his group, combining sophisticated analytic and computational approaches \cite{ref:Eberly1,ref:Eberly2}.  They were able to describe in detail the emergence of revivals as well as collapses.  The numerics in their computational approach were this time converged.  The new feature they showed, that being the revivals of the atomic inversion, are important as they showed how sensitive the atom-field coupling is to the discreteness of photon numbers. They showed the revival time is given by

\begin{equation}
    t_{R} = (2\pi/\lambda)\bar{n}^{1/2}k,\;\;\;\;\bar{n}\gg1,\;\;k=1,2,3,\dots\;. 
    \lab{2}{1}
\end{equation}

\begin{figure}
    \centering
    \includegraphics[width=1\linewidth,keepaspectratio]{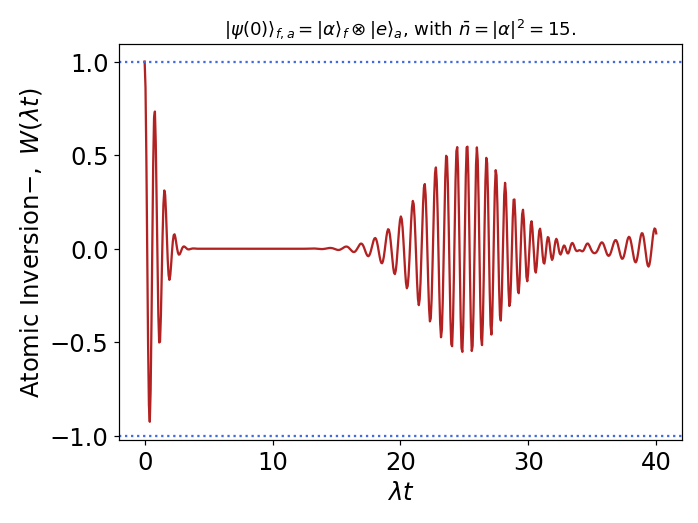}
    \caption{Atomic inversion evolution with the field prepared in an initial coherent state with $\alpha=\sqrt{15}$ and the atom initially prepared in the excited state, such that $\ket{\psi(0)}_{f,a}=\ket{\alpha,e}_{f,a}$.  Here, we plot the atomic inversion against scaled time $\lambda t$.}
    \label{fig:Atomic_Inversion}
\end{figure}

\noindent In Fig.~(\ref{fig:Atomic_Inversion}) we plot the atomic inversion for an initial coherent state with average photon number $\bar{n}=15$ as a function of scaled time $T=\lambda t$.

The relationship between the semi-classical Rabi formula for resonant excitation and the Jaynes-Cummings model is worth mention.  One might naively expect that as the photon number increases, some kind of semi-classical limit may be approached, yet this does not happen.  Knight and Radmore \cite{ref:Radmore} addressed this; for a cavity field prepared in a coherent state, one can transform the Hamiltonian by an inverse Glauber transformation \cite{ref:KnightBook} without approximation to one where the cavity is occupied by a classical field of appropriate amplitude but where the atom is also coupled to a vacuum field.  As the classical field drives the atom to an excited state, the atom, through quantum coupling to the field creation operator, can generate a cascade of quantum generated photon numbers: the first step in this cascade is the one-photon creation step and for that reason the collapse is governed by the inverse of the one-photon Rabi frequency.  
\black{Fleischhauer and Schleich \cite{ref:Fleischhauer_Schleich:1993} have  shown how the  Poisson summation formula provides insight into the nature of the revivals.}

\section{\label{sec:Section2a} Squeezed states}

\begin{figure}
    \centering
    \includegraphics[width=1\linewidth,keepaspectratio]{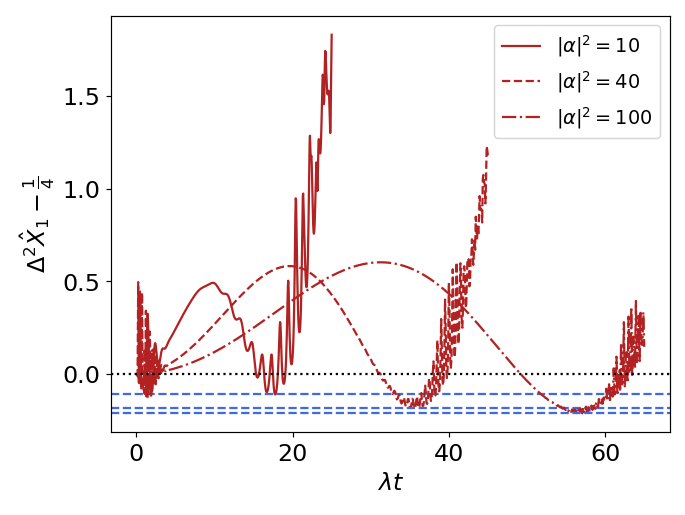}
    \caption{Quadrature squeezing with an initial coherent state with $\bar{n}=10,40,100$. We plot only the variance for the the $\hat{X}_{1}$ quadrature noting that no squeezing occurs in $\hat{X}_{2}$ at any time. A high degree of squeezing can be seen at later times for succesively higher $\bar{n}$. We include the minimum squeezing values for each case; these are: $-0.11,-0.185,\;\text{and}\;-0.213$, for increasing average photon numbers.}
    \label{fig:quadratures}
\end{figure}

Shortly after the appearance of the paper by Naroznhy \textit{et al.} \cite{ref:Eberly1} announcing the collapse and revival phenomena in the Jaynes-Cummings model, Meystre and Zubairy \cite{ref:MeystreZubairy} studied the possible occurance of quadrature squeezing in the evolution of the model with the field and the atom being at exact resonance assuming the the initial state of the field to be a coherent state of average photon number $\bar{n}(0)<10$ (this was in the days before squeezing was observed in the laboratory.) These authors found that, indeed, quadrature squeezing could occur during the evolution, but that the amount of squeezing obtained was rather modest; being about 20\% of the amount allowable below the level of the vacuum noise.  However, Kukli\'{n}ski and Madajczyk \cite{ref:Kuklinski} found that with a much greater average photon number, such as $\bar{n}(0)\sim 100$, strong squeezing can be obtained even though they included field dampening in their calculations.  These results are demonstrated in Fig.~(\ref{fig:quadratures}), where the usual quadrature operators are given by 

\begin{equation}
    \hat{X}_{1} = \frac{1}{2}\left(\hat{a}+\hat{a}^\dagger\right),\;\;\;\;\;\;\hat{X}_{2}=\frac{i}{2}\left(\hat{a}-\hat{a}^\dagger\right),
    \lab{3}{1}
\end{equation}

\noindent and we plot $s_{1}=\Delta^{2}\hat{X}_{1}-\tfrac{1}{4}$, where $\Delta^{2}\hat{X}_{1}=\braket{\hat{X}_{1}^{2}}-\braket{\hat{X}_{1}}^{2}$ is the quadrature variance and squeezing is observed for values $s_{1}<0$ with the lower bound of $s^{(\text{min.})}_{1}=-1/4$. Interestingly, they found that squeezing can be even stronger for the out-of-resonance case.  Hillery \cite{ref:Hillery1} subsequently showed for small initial photon number, the squeezing obtained in this model will always be weak.  He also pointed out that for some quantum effects to be substantial, large photon numbers are required.  

\begin{figure}
    \centering
    \includegraphics[width=1\linewidth,keepaspectratio]{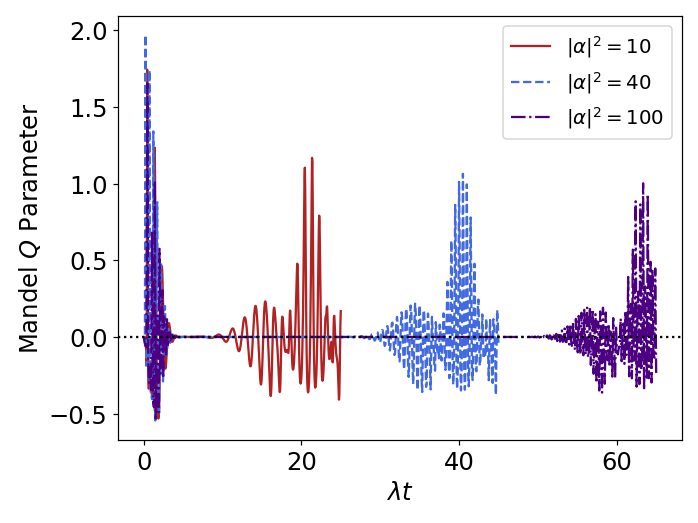}
    \caption{Mandel $Q$ parameter for different intensities of the initial field $\bar{n}=10,40,100$.  Note that for all initial coherent (Poissonian) field intensities the state displays both super-, and sub-Poissonian statistics at some point in the evolution. Note, the initial collapse time is insensitive to the photon number $\bar{n}$, while the subsequent revival times are. The revival of the varying statistics occurs at shorter times for smaller average photon numbers. }
    \label{fig:Mandel_Q}
\end{figure}

For the same cases considered in Fig.~(\ref{fig:quadratures}), we display the corresponding Mandel $Q$ parameter defined as 

\begin{equation}
    Q = \frac{\Delta^{2}\hat{n}}{\braket{\hat{n}}} - 1,
    \lab{3}{2}
\end{equation}

\noindent where $\Delta^{2}\hat{n}$ represents the photon number fluctuations.  For states displaying Poissonian statistics, the photon number fluctuations are equal to the average photon number and we have $Q=0$.  For $Q>0$ and $Q<0$, the state is said to have super$-$, and sub-Poissonian statistics, respectively (the former having large photon number variances, an example of which being thermal states, while the latter has comparably small variances like number states, and is a strictly non-classical characteristic). We plot the Mandel $Q$ parameter in Fig.~(\ref{fig:Mandel_Q}) demonstrating variation in the field photon statistics throughout the state evolution.

Alsing, Guo and Carmichael \cite{ref:Alsing} calculated the quasienergies and steady states of a coupled two-level atom and quantized electromagnetic cavity mode with the cavity mode driven by a periodic classical field that are all on resonance.  The quasienergies resulted in shifted Jaynes-Cummings level splittings which were reduced by the interaction with the driving field and vanished at a threshold value of the driving field strength.  Above this threshold, the discrete quasienergies and normalizable steady states do not exist.  For weak driving fields below the threshold, the steady states were squeezed and displaced Jaynes-Cummings eigenstates which gave rise to squeezing-induced linewidth narrowing in the vacuum Rabi splitting for the ground state of the driven Jaynes-Cummings system.  

\begin{figure*}
    \centering
    \vspace{1cm}
    \includegraphics[width=0.9\linewidth,keepaspectratio]{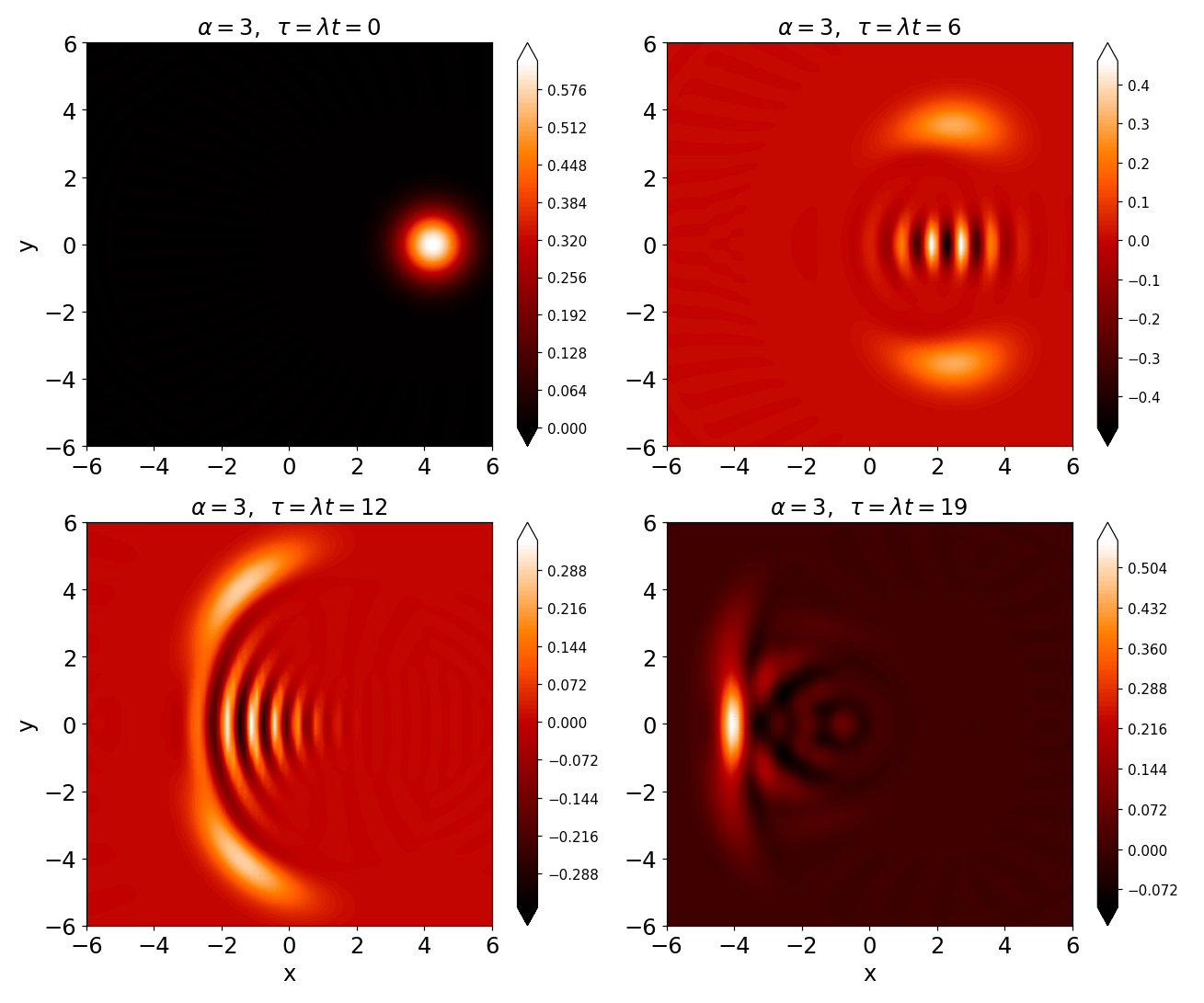}
    \caption{Evolution of the Wigner function for an initial coherent state with $|\alpha|^{2}=9$. The distribution bifurcates and counter propagates in phase space before recombining.  When the two localized distributions are opposite each other, the field is at its most cat-like. Note the negativity in the Wigner function for scaled times $\lambda t >0$ throughout the evolution suggesting non-classicality of the field.}
    \label{fig:Wigner}
\end{figure*}

In the cases mentioned in the previous paragraph, only continuous evolution was involved, with atomic and field states generally entangled.  However, as shown by Gerry and Ghosh \cite{ref:GerryGhosh}, it is possible to obtain a greater degree of squeezing, even in the case of low photon number, provided selective measurements are performed on the atom.  The idea is that an atom passes through the cavity and interacts with cavity field, then leaves the cavity where it can be subjected to classical radiation fields which implement what amounts to Ramsey pulses, after which the state of the atom is detected by selective ionization.  This process projects the cavity field into a pure state, and it was found in \cite{ref:GerryGhosh} that squeezing up to 75\% could be attained in this matter.  This procedure for projecting a cavity field into a pure state with non-classical features, such as cat states of the form $\ket{\psi_{\pm}}\propto\ket{\alpha}\pm\ket{-\alpha}$ has long been discussed and perfected in the laboratory by Haroche and collaborators in the context of the dispersive model of atom-field interactions (see \cite{ref:HarocheBook}), but applied here to the resonant case.  The possible use of these techniques in the resonant model was first proposed by Zagury and de Toledo Piza \cite{ref:Zagury} who showed that large-scale correlation effects could be obtained by the pre-selection and post-selection of atomic states.  Ghost and Gerry \cite{ref:GhoshGerry} later investigated the occurence of sub-Poissonian statistics in the field by a sequence of properly prepared and detected atoms.  We note there has been recent activity in this area \cite{ref:Moya,ref:Bohman}.

\begin{figure}
    \centering
    \includegraphics[width=1\linewidth,keepaspectratio]{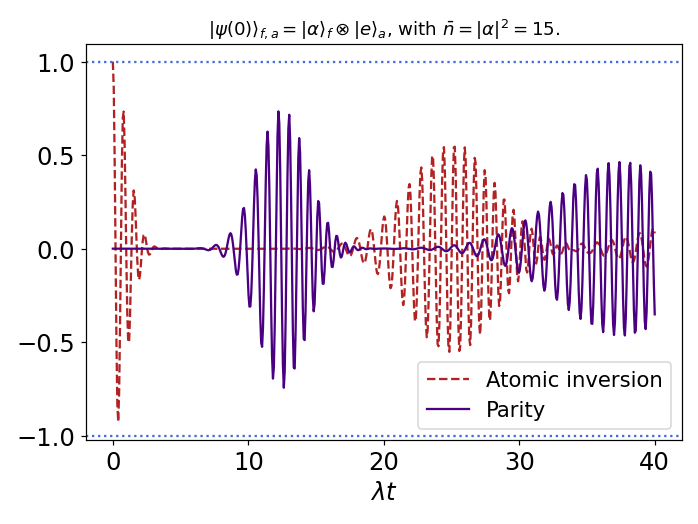}
    \caption{Expectation value of the field parity operator (purple, dashed) plotted against scaled time $\lambda t$ for the initial conditions $\alpha=\sqrt{15}$ and the atom initially prepared in the excited state, such that $\ket{\psi(0)}_{f,a}=\ket{\alpha,e}_{f,a}$.  We include a plot of the atomic inversion (red,solid) for reference}
    \label{fig:Parity}
\end{figure}

The full range of the occurence of non-classical field effects in the Jaynes-Cummings model can be found in the recent book by Larson and Mavrogordaros \cite{ref:Larson}.

\section{\label{sec:Section3} Atom-Field Schr\"{o}dinger Cats}

Further understanding can be obtained by studying the evolution of the field state in phase space using quasi-probabilities.  A single field mode in the JCM interacting with a single two-level atom exhibits a two-component behaviour easily described by the Schmidt basis \cite{ref:Phoenix,ref:Gea,ref:Ekert}. A field initially in a coherent state has a Gaussian Wigner function which evolves by bifurcating into two counter-rotating components; the point in time at which the two components are maximally separated coincides with the collapse time.  Then, when they come to a point 180 degrees from the starting point, we see a revival. This is demonstrated in Fig.~(\ref{fig:Wigner}) for an initial coherent state with $\alpha=3$.  The JCM bifurcation was noted in early papers by Eiselt and Risken \cite{ref:Eiselt1,ref:Eiselt2} who showed that the bifurcation could be seen in the evolution of both the $Q$-function and Wigner function of the field.

At the point of maximal separation, which occurs at the midpoint between the collapse and first revival of the Rabi oscillations, the atom-field system becomes separable as was shown by Knight and Phoenix \cite{ref:Phoenix} and by Gea Banacloche \cite{ref:Gea}. The field state at this time is approximately described by \black{(unnormalized)}
Schr\"{o}dinger cat states of the form \cite{ref:GerryAJP}

\begin{equation}
    \ket{\psi_{\pm}}\black{\sim}\ket{\beta}\pm\ket{-\beta},
    \lab{4}{1}
\end{equation}

\noindent where $\ket{\pm\beta}$ are coherent states. Note that $\ket{\psi_{+}}$ contains only even photon numbers while $\ket{\psi_{-}}$ contains only odd.  We mention this because even though the atom and field states are approximately separable at the aforementioned midpoint, the field is not static.  That this is so becomes apparent if one calculates the expectation value of the field photon number parity operator, given by $\hat{\Pi}=(-1)^{\hat{a}^\dagger\hat{a}}$, for the resonant JCM as was done by Birrittella \textit{et al.} \cite{ref:Birrittella}.  The result is displayed in Fig.~(\ref{fig:Parity}), where it is evident that at the midpoint of the period where the Rabi oscillations of the atomic inversion are quiescent, the photon number parity is rapidly oscillating.  This indicates that the field state at these times is rapidly switching between states of even and odd parity, i.e. switching between even and odd cat-like states.  Ultimately, this can be attributed to quantum interference efffects between the widely separated counter-rotating components of the field. These oscillations in the photon number parity have been seen in a recent experiment by the Haroche group \cite{ref:Haroche}.

But what happens when instead of an entirely closed loss-free system, we allow dissipation? Barnett and Knight \cite{ref:Barnett} were first to show revivals were very sensitive to dissipation; collapses less so.

\section{\label{sec:Section4} Experimental observations}

The experimental investigations of the JCM came out of work on the radiative properties of highly excited Rydberg atoms in cavity QED.  Atoms prepared in a very highly excited states have huge dipole moments, and combined with cavities with very high quality factors allow the study of atom-field couplings at low photon number (Rempe \textit{et al.} \cite{ref:Rempe}, Haroche and Raimond \cite{ref:HarocheBook}), resulting in clear observations of both collapses and revivals of the Rabi oscillations.  But this is not the only way to realize a spin-boson JCM Hamiltonian: another way uses laser-cooled trapped ions.  A laser cooled ion in an electromagnetic trap, such as an RF Paul trap \cite{ref:Paul} realizes the Jaynes-Cummings interaction between internal states of the ion and the quantized harmonic vibrational motion of the center-of-mass (CM) of the ion which is constrained by the trap.  The Jaynes-Cummings interaction is engineered by the use of a tunable laser field directed along the axis of the trap. If $\nu$ is the angular frequency of the vibrational motion of the CM of the ion, which is determined by the trap parameters, and if $\omega_{0}$ is the angular frequency associated with two relevant internal states of the ion, then with the laser frequency tuned such that $\omega_{L}=\omega_{0}+\nu$, one ends up with an interaction that contains, after discarding rapidly oscillation terms, the Jaynes-Cummings interaction $-i\hat{a}^\dagger\hat{\sigma}_{-}+h.c.$ wherein the Bose operators $\hat{a},\hat{a}^\dagger$ are understood to represent the quantized vibrational motion of the CM of the ion.   This replaces the quantized single-mode field: i.e. we have phonons instead of photons.  Indeed, with the choice $\omega_{L}=\omega_{0}-\nu$, the otherwise unrealizeable ``anti-Jaynes-Cummings" interaction $-i\hat{a}^\dagger\hat{\sigma}_{-}+h.c.$ can be engineered.

As far as we are aware, the first laboratory realization of the JCM in the context of trapped ions was reported by Meekhof \textit{et al.} \cite{ref:Meekhof}.  With the vibrational motion of the ion prepared in coherent state by implementing a displacement of the ion trap itself with the ion initially in its vibrational ground state, Meekhoff \textit{et al.} realized the coherent state JCM.  Clear evidence of the predicated collapse and revival of the Rabi oscillations is presented in \cite{ref:Meekhof}.

A significant physical implementation and translation of the JCM beyond the field of quantum optics has been to circuit QED systems consisting of a single superconducting qubit coupled to the electromagnetic field of a single mode inside a microwave resonator (see the review article by Schmidt and Koch \cite{ref:Koch} and references therein, with comparison/contrast to cavity QED systems).

\section{\label{sec:Section5} Closing Remarks}

The Jaynes-Cummings model continues to be one of the key conceptual models of current quantum optics: it continues to be very widely cited with no diminution of its perceived importance.

In closing, it may seem ironic that E. T. Jaynes, passionate about looking for ways to avoid field quantization \black{\cite{ref:Jaynes:1978}}, was the creator of the most fundamental model of atom-photon interaction; one widely adopted and confirmed by experiment.  Extensive reviews of the JCM exist, for example Shore and Knight \cite{ref:Shore}, and more recently and exhaustively by Larson and Mavrogordatos \cite{ref:Larson}, indicating its lasting importance.  

In the 60+ years since the publication of the Jaynes-Cummings model, the paper remains a seminal contribution to quantum optics, cited many times each year.  To the current authors, the key contribution is a demonstration of a measurable effect that showed the discreteness of the quantized radiation field.  The Cummings collapse reflects the contribution of incommensurate coupling strengths; of course any distribution of field strengths will lead to dephasing analogous to a collapse, but the revival is truly dependent on the photon number being discrete.  This will happen even with fields showing little sign otherwise of ``non-classicality" including coherent and thermal fields.  To us, this is really the `take-home' message of the Jaynes-Cummings model.

\section{Acknowledgements}
 Any opinions, findings and conclusions or recommendations expressed in this material are those of the author(s) and do not necessarily reflect the views of their home institutions.
 

\bibliography{JC_bib}

\end{document}